\documentclass[aip,preprint]{revtex4-1}

\usepackage{graphicx}
\usepackage{graphics}
\usepackage{epsfig}
\usepackage{ulem}
\usepackage{color}
\usepackage{subfigure}

\begin{document}


\title{Spider-Web Inspired Mechanical Metamaterials}



\author{Marco Miniaci}
\affiliation{University of Le Havre, Laboratoire Ondes et Milieux Complexes, UMR CNRS 6294, 75 Rue Bellot, 76600 Le Havre, France}

\author{Anastasiia Krushynska}
\affiliation{Department of Physics, University of Torino, Via Pietro Giuria 1, 10125, Torino, Italy}

\author{Alexander B. Movchan}
\affiliation{Department of Mathematical Sciences, University of Liverpool, L69 3BX, Liverpool, UK}

\author{Federico Bosia}
\affiliation{Department of Physics, University of Torino, Via Pietro Giuria 1, 10125, Torino, Italy}

\author{Nicola M. Pugno}
\email[Corresponding author:]{ nicola.pugno@unitn.it}
\affiliation{Laboratory of Bio-Inspired \& Graphene Nanomechanics, Department of Civil, Environmental and Mechanical Engineering, University of Trento, Via Mesiano, 77, 38123, Trento, Italy}
\affiliation{Center for Materials and Microsystems, Fondazione Bruno Kessler, Via Sommarive 18, 38123, Povo (Trento), Italy}
\affiliation{School of Engineering \& Materials Science, Queen Mary University of London, Mile End Road, London, E1 4NS, UK}


\date{\today}

\begin{abstract}

Spider silk is a remarkable example of bio-material with superior mechanical characteristics. Its multilevel structural organization of dragline and viscid silk leads to unusual and tunable properties, extensively studied from a quasi-static point of view. In this study, inspired by the Nephila spider orb web architecture, we propose a novel design for mechanical metamaterials based on its periodic repetition.
We demonstrate that spider-web metamaterial structure plays an important  role in the dynamic response and wave attenuation mechanisms. The capability of the resulting structure to inhibit elastic wave propagation in sub-wavelength frequency ranges is assessed and parametric studies are performed to derive optimal configurations and constituent mechanical properties. The results show promise for the design of innovative lightweight structures for tunable vibration damping and impact protection, or the protection of large scale infrastructure such as suspended bridges.

\end{abstract}

\pacs{}

\maketitle 


Many natural materials display outstanding properties that can be attributed to 
their complex structural design, developed in the course of millions of years of evolution~\cite{Gao2003, Aizenberg2005, Kamat2000}. 
Particularly fascinating are spider silks, which exhibit unrivaled strength and toughness 
when compared to most materials~\cite{Vollrath1992, Gosline1999, Boutry2009, Meyer2014}. Previous studies have revealed 
that mechanical performance of spider webs is not only due to the remarkable properties of the silk material, but 
also to an optimized architecture that is adapted to different functions~\cite{Cranford2012, Zaera2014}.

Structural behaviour of orb spider webs has been extensively analyzed under 
quasi-static~\cite{Boutry2009, Cranford2012, Aoyanagi2010} and
dynamic~\cite{Alam2007, Ko2004} loading conditions. However, 
the spider web structure has yet to be exploited for the design of
mechanical metamaterials. 
The latter include phononic crystals and elastic metamaterials, and 
are usually periodic composites capable of inhibiting the propagation of elastic or 
acoustic waves in specific frequency ranges called band gaps.
This unique ability opens a wide range of application opportunities, such as 
seismic wave insulation~\cite{Miniaci2016a}, 
environmental noise reduction~\cite{MartinezSala1995}, 
sub-wavelength imaging and focusing~\cite{Bigoni2013},
acoustic cloaking~\cite{Farhat2008},
and even thermal control~\cite{Maldovan2013}.
In phononic crystals, band gaps are induced due to Bragg scattering 
from periodic inhomogeneities~\cite{Brillouin1946},
while in elastic metamaterials, sub-wavelength band gaps 
can be generated due to localized resonances~\cite{Liu2000}.
The latter are commonly achieved by 
employing heavy constituents~\cite{Liu2000, Pennec2010, Hussein2014, Krushynska2014}.
However, recently it has been found that hierarchically organized~\cite{Miniaci2016b} or lattice-type structures can also generate sub-wavelength band gaps~\cite{LimBertoldi2015, WangBertoldi2015}. From this perspective, a spider-web inspired, lattice-based elastic metamaterial 
seems to be a promising alternative to control low-frequency wave propagation.

In this letter, we design a novel metamaterial 
inspired by the Nephila orb web architecture and analyze the dynamics of elastic 
waves propagating in the considered structures. The capability of inducing local resonance band gaps is studied by highlighting
the importance of the structural topology as well as the material properties of single constituents. 
We find that the band gaps are associated with either ring-shaped resonators or constituent parts of the bearing frame 
and can be tuned to desired frequencies by varying material parameters of the constituents or the number of the ring resonators. 
Overall, our analysis shows that the spider web-inspired architectures allow to extend the effective attenuation frequency ranges compared to a simple lattice~\cite{Martinsson2003} and simultaneously provide lightweight yet robust structures.

\begin{figure}[htb]
\centering
\subfigure[]
{\includegraphics[trim=4cm 5.5cm 11cm 1.2cm, clip=true, width=.25\textwidth]{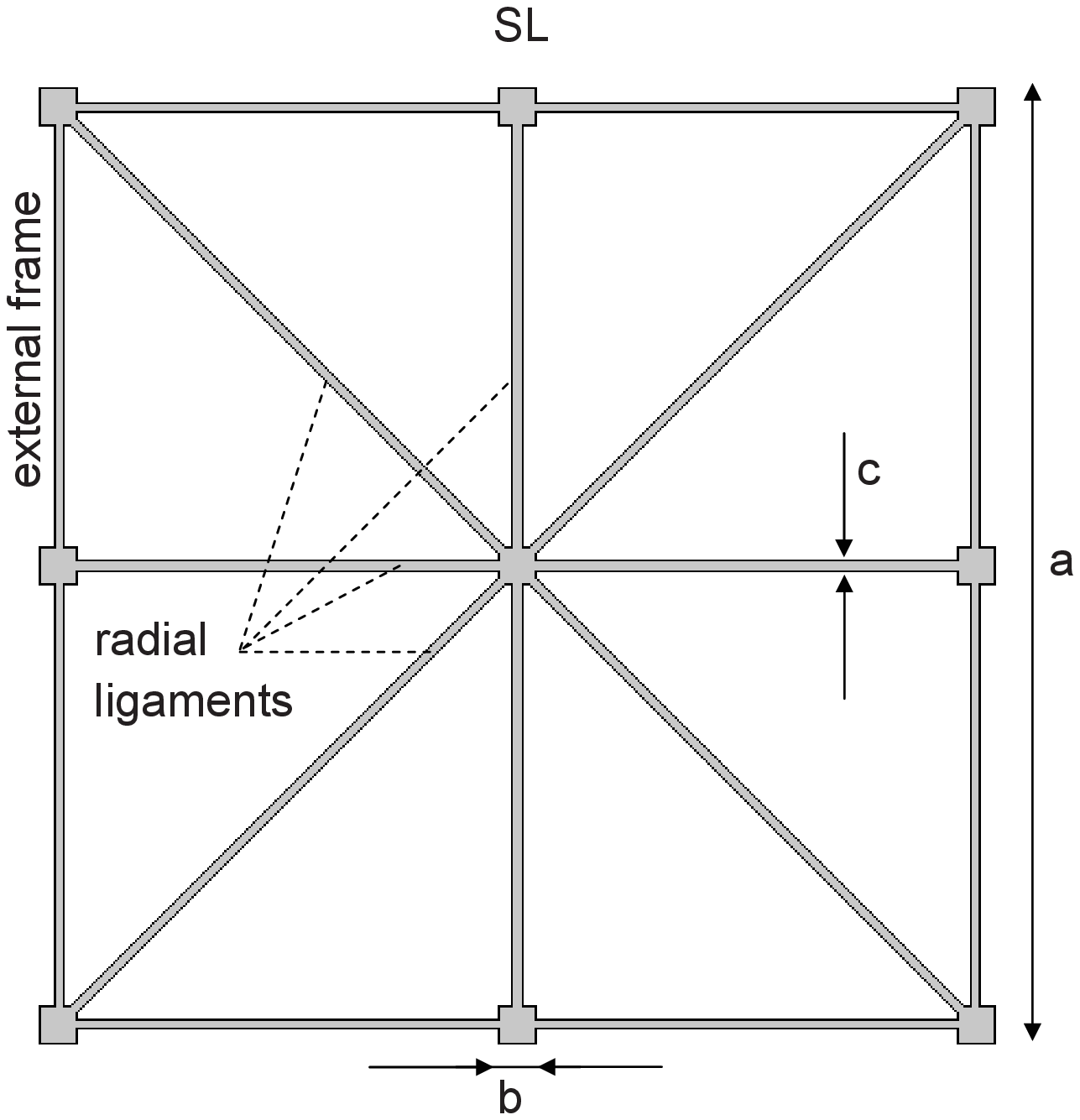}}
\subfigure[]
{\includegraphics[trim=4cm 5.5cm 11cm 1.2cm, clip=true, width=.25\textwidth]{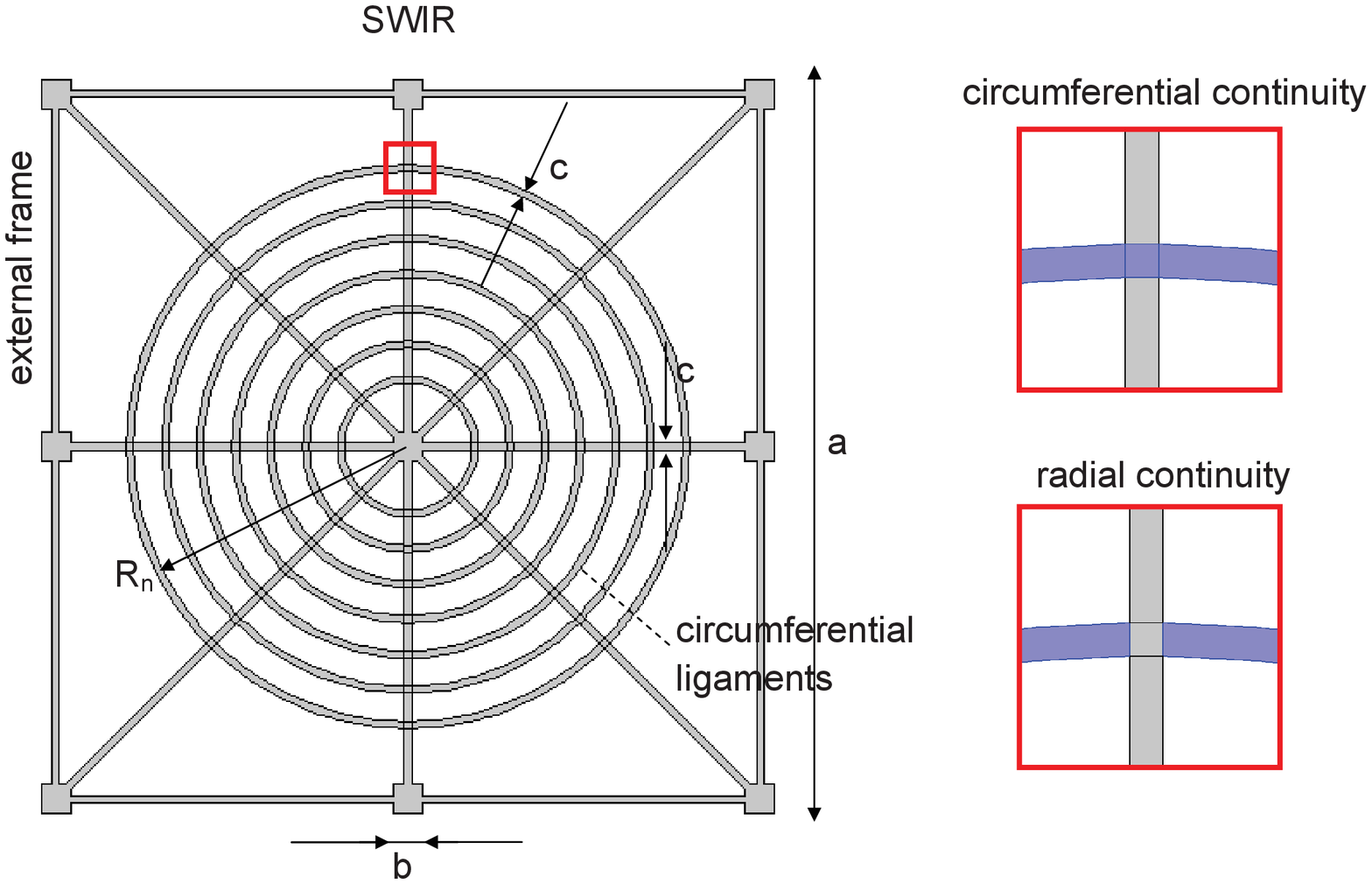}}
\caption{(a) Bearing frame and (b) spider-web inspired unit cells for lattice-based mechanical metamaterials.} 
\label{models}
\end{figure}

We consider a spider web-inspired metamaterial in the form of an infinite in-plane lattice modeled
by periodically repeating representative unit cells in a square array.
The primary framework of the unit cell is a square frame with supporting radial ligaments (Fig. \ref{models}a).
The ligaments intersect the frame at right-angle junctions acting as `hinge' joints (square junctions in Fig. \ref{models}a). 
The secondary framework is defined by a set of equidistant circumferential ligaments (or ring resonators) 
attached to the radial ligaments by hinge joints, further called `connectors' to distinguish them from 
the joints in the first framework (Fig. \ref{models}b).
The geometry of the metamaterial is completely defined by 5 parameters: 
unit cell pitch $a$, 
size of square joints $b$, 
thickness of radial and circumferential ligaments $c$, 
number of ring resonators $N$,
and 
radius of a ring resonator $R_N$.
We initially consider $a = 1 [m]$, 
$b = 0.04 \cdot a$,
$c = 0.01 \cdot a$,
$N=7$,
and
$R_N = 0.1 \cdot a \cdot (N+1)/2$.
The material properties of the primary and secondary frameworks correspond to the parameters of 
dragline ($E_{d} = 12$ GPa, $\nu_{d} = 0.4$, $\rho_{d} = 1200$ kg$/$m$^3$) and viscid ($E_{v} = 1.2$ GPa, $\nu_{v} = 0.4$, $\rho_{v} = 1200$ kg$/$m$^3$) silks of the Nephila orb spider web~\cite{Zaera2014}, respectively.
Material properties of the connectors can assume dragline or viscid silk values, as specified below.

The propagation of elastic waves in infinite and finite-size lattices is  
investigated numerically by using the Finite Element (FE) commercial package COMSOL Multiphysics~\cite{COMSOL}. 
Unit cells are discretized by triangular linear elements of size $a/20$. 
Wave dispersion in the infinite lattice
is studied by applying the Bloch conditions~\cite{Hussein2014} at the unit cell boundaries and performing 
frequency modal analysis for wavenumbers along the 3 high-symmetry directions $\Gamma-X-M$ within the first irreducible Brillouin zone \cite{Miniaci2015}. 

\begin{figure}[htb]
\centering
\subfigure[]
{\includegraphics[trim=2cm 16cm 6.3cm 2.5cm, clip=true, width=.45\textwidth]{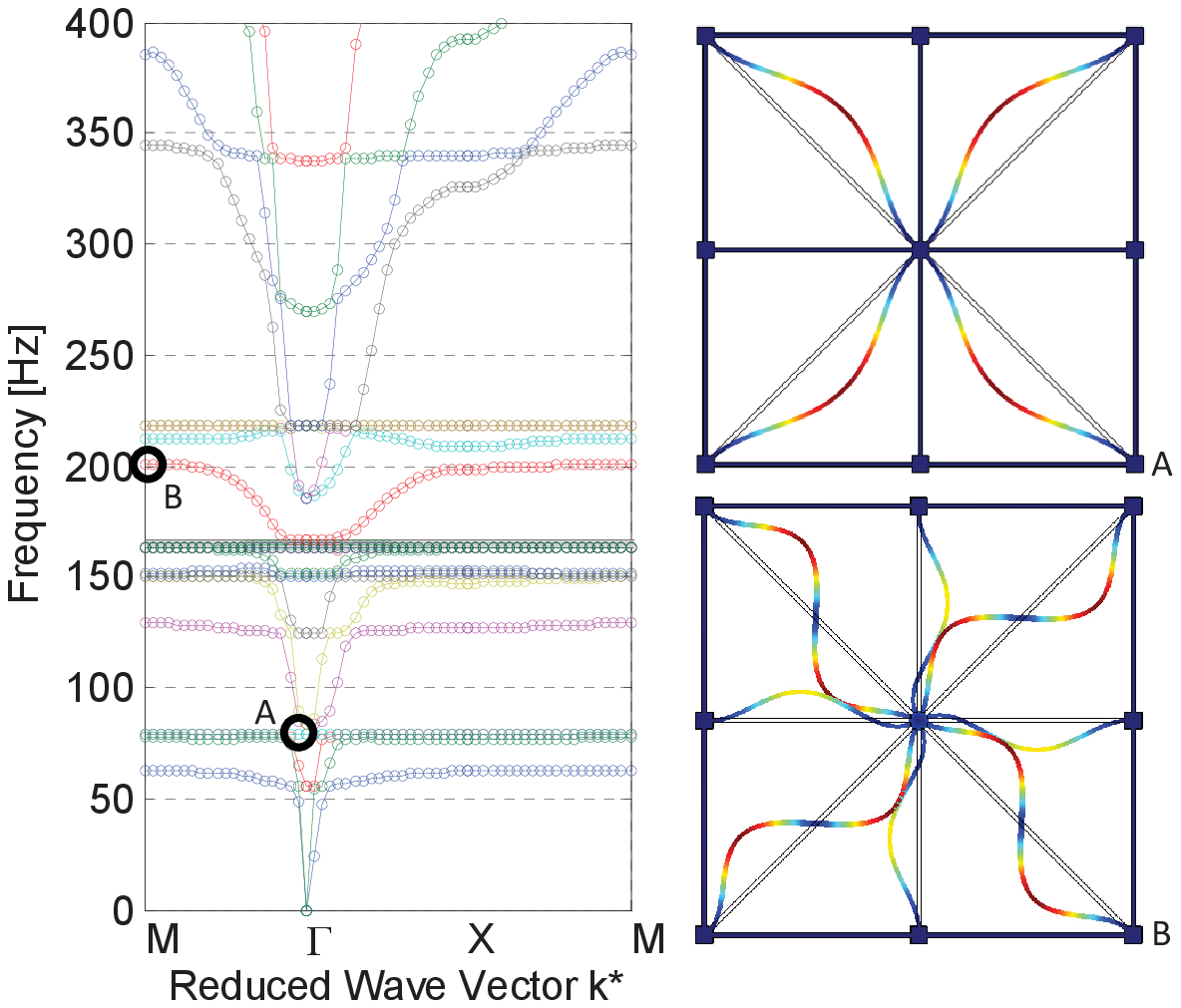}}
\subfigure[]
{\includegraphics[trim=2cm 16cm 6.3cm 2.5cm, clip=true, width=.45\textwidth]{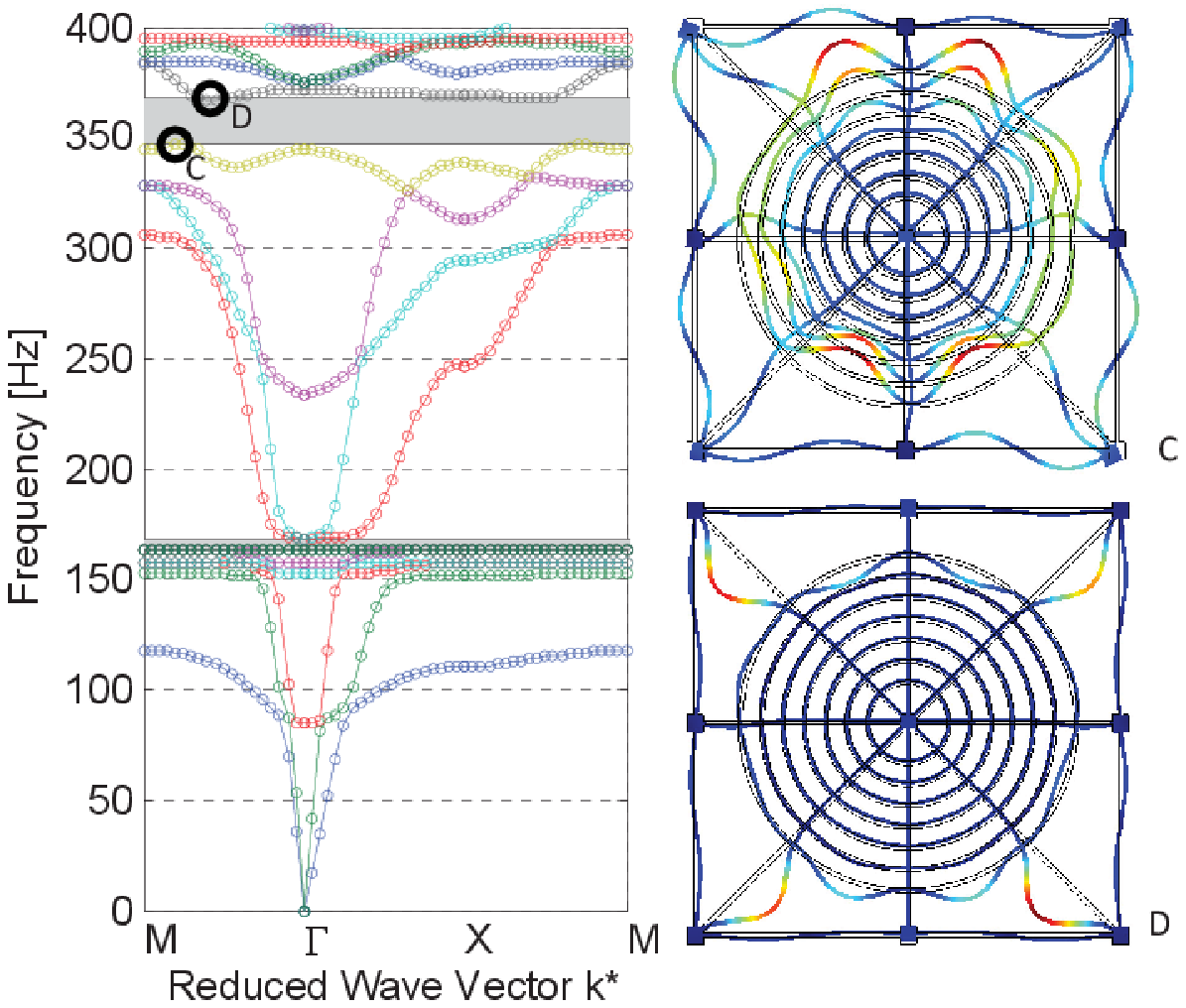}}
\subfigure[]
{\includegraphics[trim=2cm 16cm 6.3cm 2.5cm, clip=true, width=.45\textwidth]{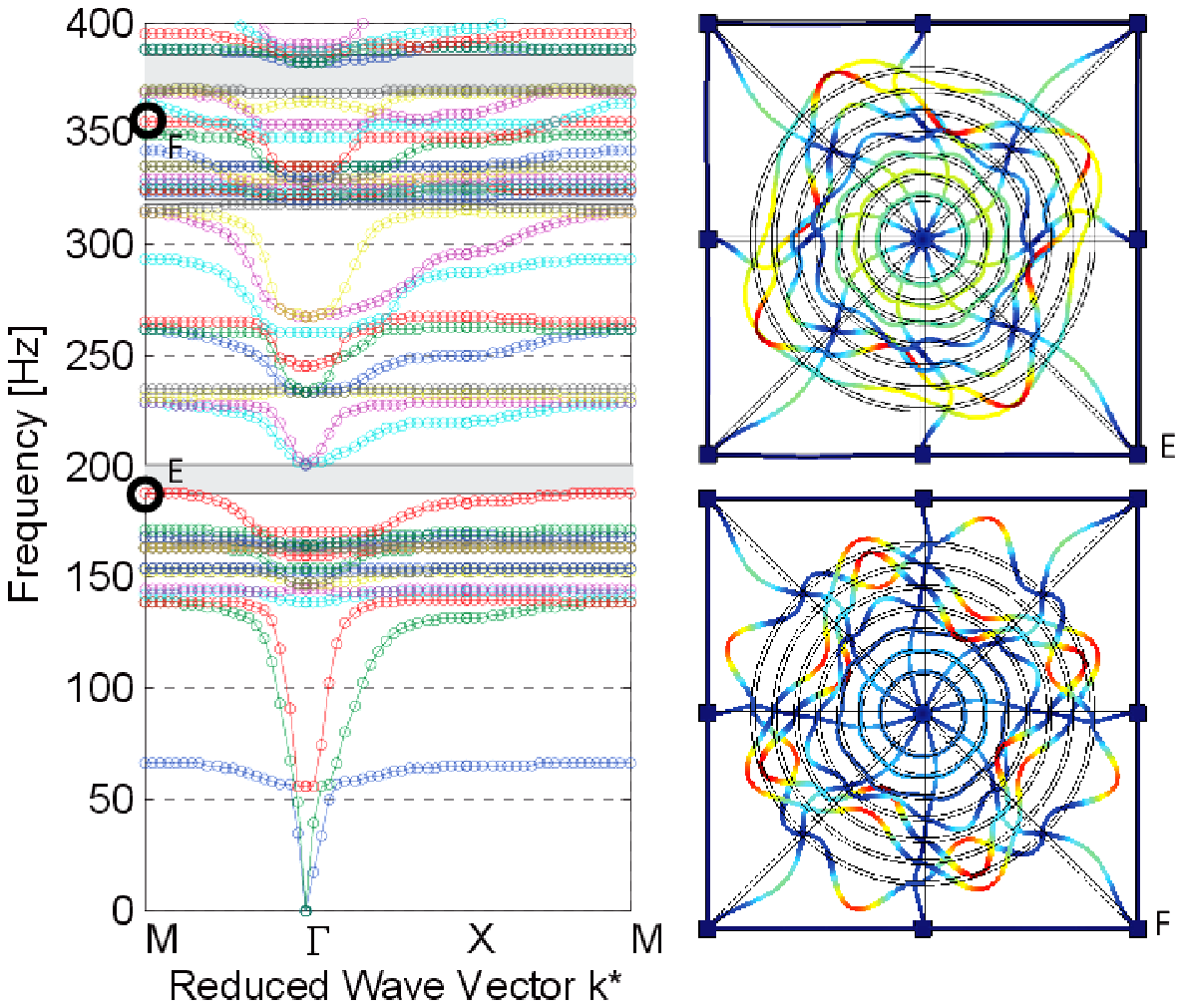}}
\subfigure[]
{\includegraphics[trim=2cm 16cm 6.3cm 2.5cm, clip=true, width=.45\textwidth]{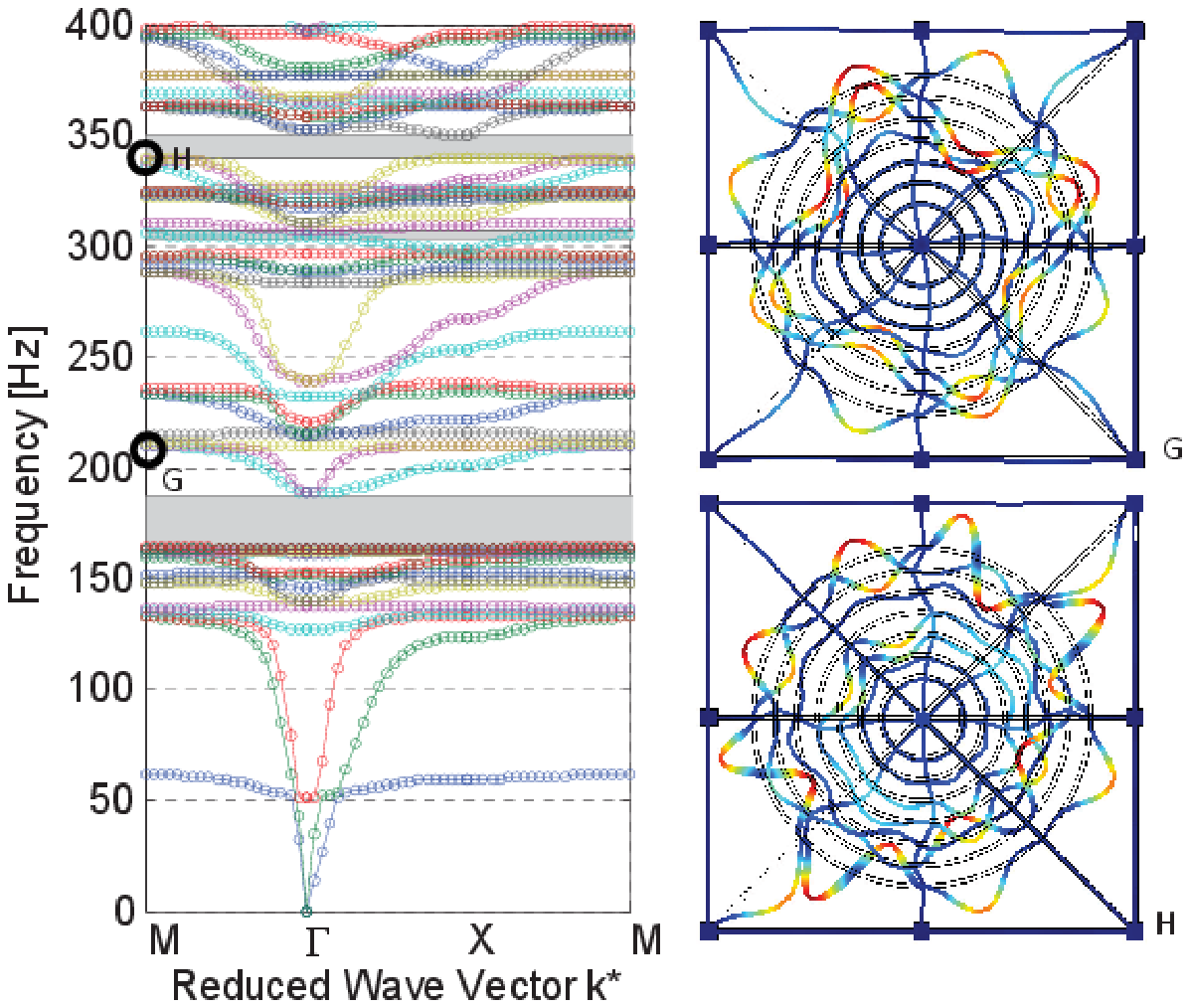}}
\caption{(a) Band structure for a regular lattice represented by the first framework. (b)-(d) Band structures for spider-web inspired lattices with a stiff frame and (b) stiff ring resonators (dragline silk) or compliant ring resonators (viscid silk) with (c) stiff and (d) compliant connectors joining the radial and circumferential ligaments. Band gaps are shaded in gray.}
\label{DispersionCurves}
\end{figure}

First, we investigate the propagation of small-amplitude elastic waves in an infinite structure formed by
the primary framework unit cell (Fig.~\ref{models}a), called ``regular lattice'' metamaterial. Figure~\ref{DispersionCurves}a shows the band structure for the regular lattice as a function of reduced wave vector $k^* = [k_x a/\pi; k_y a/\pi]$, where it appears that
there are no band gaps in the considered frequency range, up to 400 Hz. However, the structure is characterized by several localized modes at various frequencies represented by (almost) flat bands. Consideration of the corresponding vibration patterns reveals that the motions are localized within the radial ligaments. 

Next, the circumferential elements are introduced to analyze the wave dynamics in a spider-web inspired metamaterial formed by the unit cell shown in Fig.~\ref{models}b. Here, we explore three possibilities: 1) the circumferential ligaments have the same mechanical properties as the radial ligaments (dragline silk); 2) the circumferential ligaments have the mechanical properties of viscid silk, while connectors of radial and circumferential ligaments have the properties of dragline silk, and 3) both the circumferential ligaments and the connectors have the properties of viscid silk. This analysis allows to evaluate the influence of material parameters on the wave attenuation performance of the spider-web topology.

The band structure of the spider-web metamaterial made of the dragline silk material is shown in Figure~\ref{DispersionCurves}b and exhibits a complete band gap between the $10^{th}$ and $11^{th}$ bands at frequencies from $346.5$ to $367.4$ Hz, which is shaded in light gray. Vibration patterns at the band gap bounds (points C and D in Fig.~\ref{DispersionCurves}b) reveal that the whole unit cell is involved in the motion. As the band gap bounds are formed by non-flat curves, the band gap is not due to local resonance effects. At the same time, it cannot be induced by Bragg scattering, since it is located at least twice below the frequencies at which a half-wavelength of either longitudinal (2314 Hz) or shear (945 Hz) waves in the silk material is equal to the unit cell size, and the bounds are shifted from the edge points ($\Gamma$, $X$, and $M$) of the irreducible Brillouin zone. 
Further analysis of the band gap origin is beyond the scope of this letter, since we are mainly focusing on a spider-web inspired structure with \textit{different} mechanical properties for radial and circumferential ligaments~\cite{Cranford2012, Zaera2014}. Another remarkable feature of the band structure in Fig.~\ref{DispersionCurves}b is the smaller number of localized modes compared to Fig.~\ref{DispersionCurves}a, which may be explained with the elimination of local resonances due to the coupling between motions in radial and circumferential ligaments.

The assignment of the material parameters of viscid silk to ring resonators leads to the appearance of two band gaps in the corresponding band structures (Fig.~\ref{DispersionCurves}c-d) regardless of the mechanical properties of the connectors joining radial and circumferential ligaments. Due to the more compliant properties of the resonators, the band gaps are located at lower frequencies compared to the case in Fig.~\ref{DispersionCurves}b. Moreover, it is obvious that these are so-called hybridization band gaps~\cite{Sainidou2002} induced by local resonances, since the lower bounds are formed by flat curves representing localized motions (see modes relative to points E-H in Fig.~\ref{DispersionCurves}c-d), and the reduced Bloch wave vector $k^*$ has a $\pi$ change inside each band gap. In the case when the ring resonators and the connectors have the same mechanical properties, the band gaps are shifted to lower frequencies due to a more compliant behavior of the connectors (Fig.~\ref{DispersionCurves}d). 

Another peculiarity of the band structures in Fig.~\ref{DispersionCurves}c-d in comparison to Fig.~\ref{DispersionCurves}a-b is a larger number of almost flat pass bands, many of which correspond to localized modes. In the cases when the ring resonators are characterized by a low stiffness the localized modes are associated to a family of standing waves mostly dominated by high inertia of the resonators (see modes relative to points E-H in Fig.~\ref{DispersionCurves}c-d). If the connectors between radial and circumferential ligaments have the same material properties as the resonators (which is the closest configuration to a real spider web), it appears that in the low-frequency regime, frequencies of standing waves may be associated with the ring resonators only. The natural frequencies $\omega_n$ for non-axisymmetric in-plane flexural vibrations of these resonators can be expressed~\cite{Tim, Love} in closed form:
\begin{equation}
\omega_n = k ~ \frac{n (n^2-1)}{R^2 \sqrt{n^2 + 1}}, ~ n > 1. \label{omega}
\end{equation}
Here $R$ stands for the radius of a ring resonator, and $k$ is a dimensional constant that depends on the elastic modulus of the ring resonator, the mass density,  and its cross-section. Examples of vibrational modes for several values of $n$ are shown in the Supplementary Material.  However, solution~(\ref{omega}) cannot be applied to a spider-web lattice system, since the dynamic response of the latter is governed by the entire structure and not the individual decoupled resonators (see the Supplementary Material for details). 

Wave propagation in the regular and spider-web lattice metamaterials  
can be better understood by analyzing the transformations of individual modes in band structures of Fig.~\ref{DispersionCurves} for varying geometrical and mechanical parameters of the unit cell. 
For example, the frequency range of the lowest (localized) mode is the same for 
the regular and spider-web-type metamaterials, except for the case in Fig.~\ref{DispersionCurves}b, 
whereas the second mode (second curve from the bottom) is shifted towards higher frequencies, 
as circumferential ligaments are introduced. These features can be explained by 
examining the vibration modes of the corresponding systems, a detailed analysis of which is given in the 
Supplementary Material.

\begin{figure}[htb]
\centering

\subfigure[]
{\includegraphics[trim=0cm 0cm 0cm 0cm, clip=true, width=.45\textwidth]{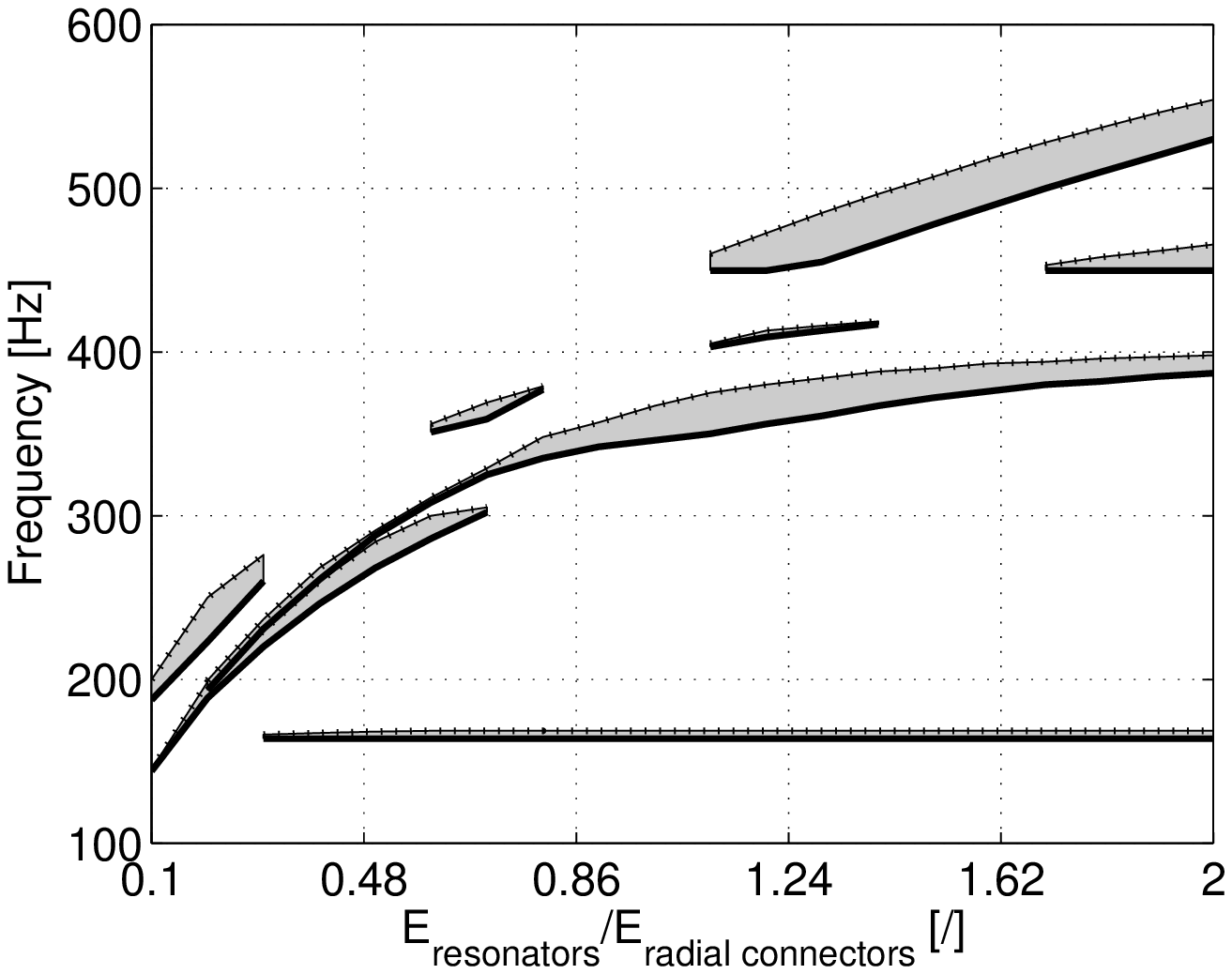}}
\subfigure[]
{\includegraphics[trim=0cm 0cm 0cm 0cm, clip=true, width=.45\textwidth]{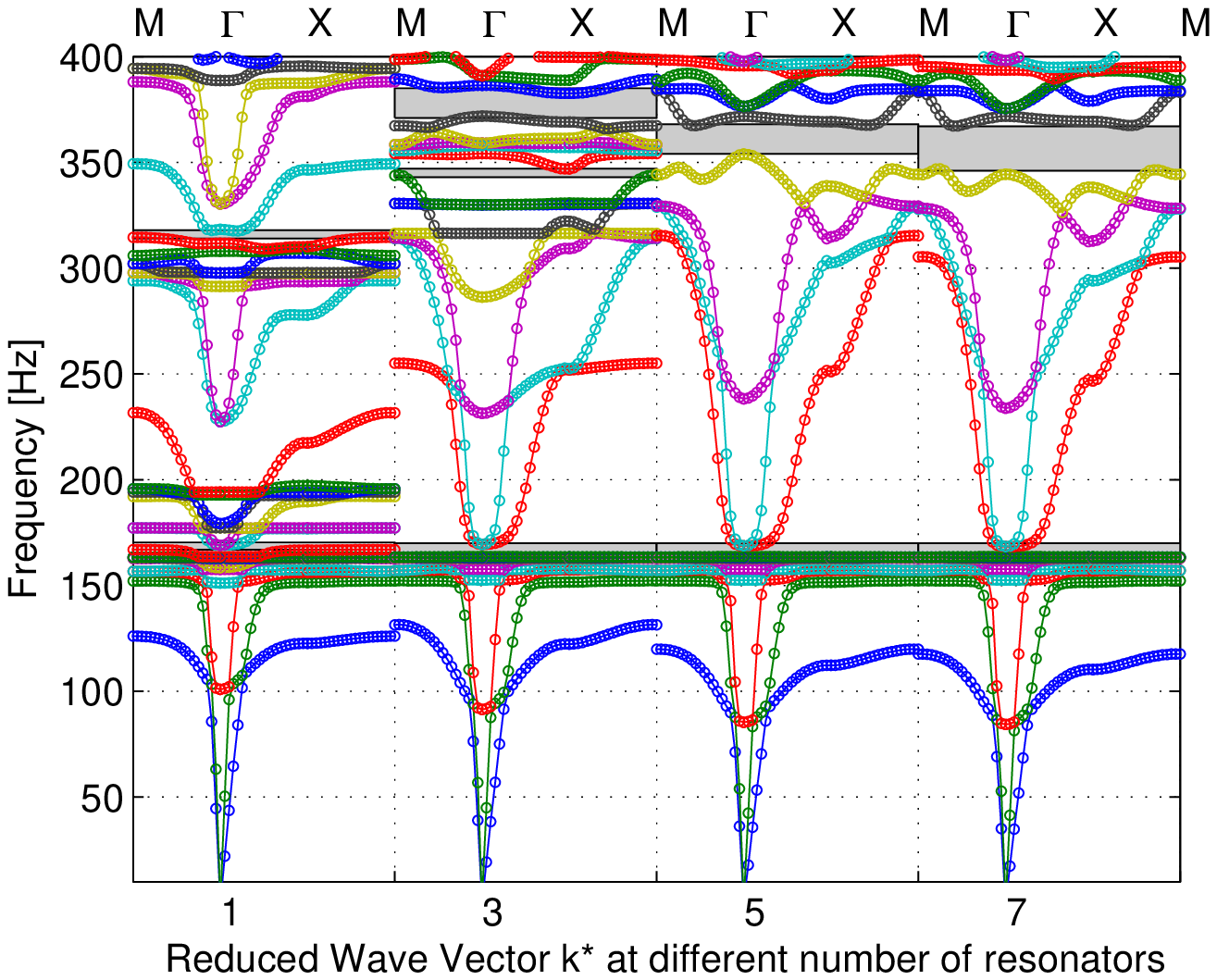}}

\caption{(a) Band gap frequencies for the lowest band gaps as functions of ratio $E_{rr}/E_{rl}$. (b) Band structures for the spider-web inspired unit cells comprising 1, 3, 5, and 7 most external ring resonators.}
\label{ParametricAnalysis}
\end{figure}

Further, we investigate the evolution of the band structure for the spider-web inspired metamaterials 
by varying the mechanical properties and number of ring resonators.
The parametric study is performed for metamaterials with the ring resonators with
intermediate stiffness values between those of dragline and viscid silks.  
The overall band structure resembles that shown in Figs.~\ref{DispersionCurves}c-d. Here, we focus our attention only on the band gap frequencies.
Figure \ref{ParametricAnalysis}a shows band gap frequencies versus ratios $E_{rr}/E_{rl}$, where $E_{rr}$ and $E_{rl}$ are the stiffnesses 
of the ring resonators and radial ligaments, respectively. 
It appears that for $E_{rr}/E_{rl} = 0.6$ the first band gap is maximized, and the second one is closed. 
In general, as the stiffness of the ring resonators increases, inhibited frequency ranges are translated towards higher frequencies, except 
one narrow band gap located slightly above 150 Hz, whose frequencies are independent of the mechanical parameters of the resonators. 

Next, we investigate the wave dispersion in structures with a varying number of ring-shaped resonators. 
Figure~\ref{ParametricAnalysis}b shows dispersion diagrams for the unit cells with 1,3,5, and 7 
stiff ring resonators (dragline silk). In all cases, the unit cells with a reduced number of 
resonators are obtained by removing the most internal resonators from the original geometry shown (Fig.~\ref{models}b). 
Stiff rings are chosen here, since the corresponding band structures have
fewer pass bands, thus facilitating the analysis. In the considered frequency range, the number of localized modes decreases with the increase of the number of the ring resonators, since the structure becomes stiffer. 
This results in an increase in the number of band gaps, which are also shifted to lower frequencies, since the wave attenuation is enhanced due to increased number of resonators. Interestingly, the unit cells with 3-7 ring resonators possess
the first (narrow) band gap at the same frequencies, slightly above 150 Hz. Examination of the corresponding vibration forms reveals
that this band gap is induced by local resonances in the external square frame. Thus the gap frequencies are independent of the number of the ring resonators, as well as of their mechanical properties (Fig.~\ref{ParametricAnalysis}a).

\begin{figure}[htb]
\centering

\subfigure[]
{\includegraphics[trim=2.5cm 10cm 2.5cm 3.5cm, clip=true, width=.45\textwidth]{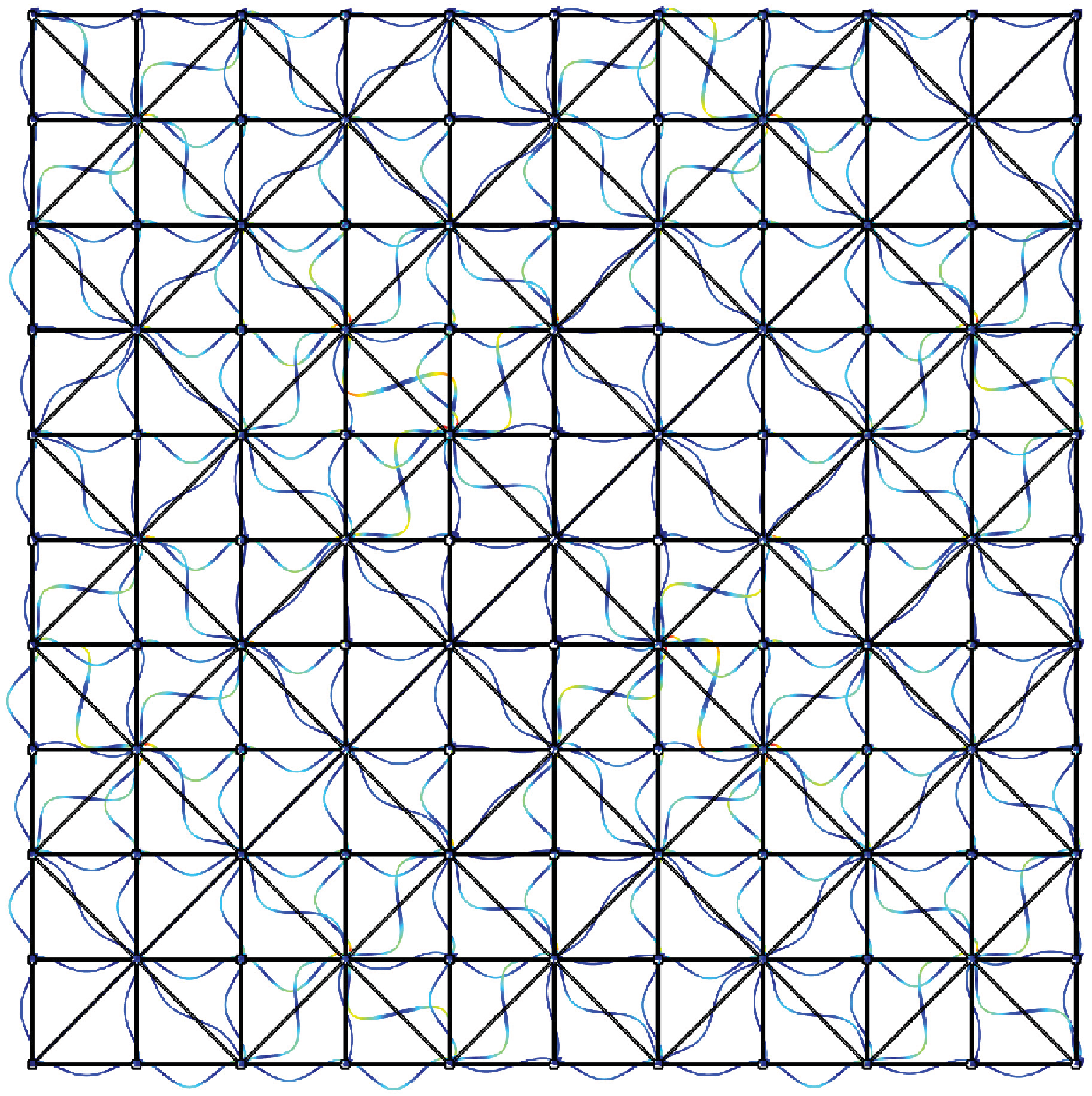}}
\subfigure[]
{\includegraphics[trim=2.5cm 10cm 2.5cm 3.5cm, clip=true, width=.45\textwidth]{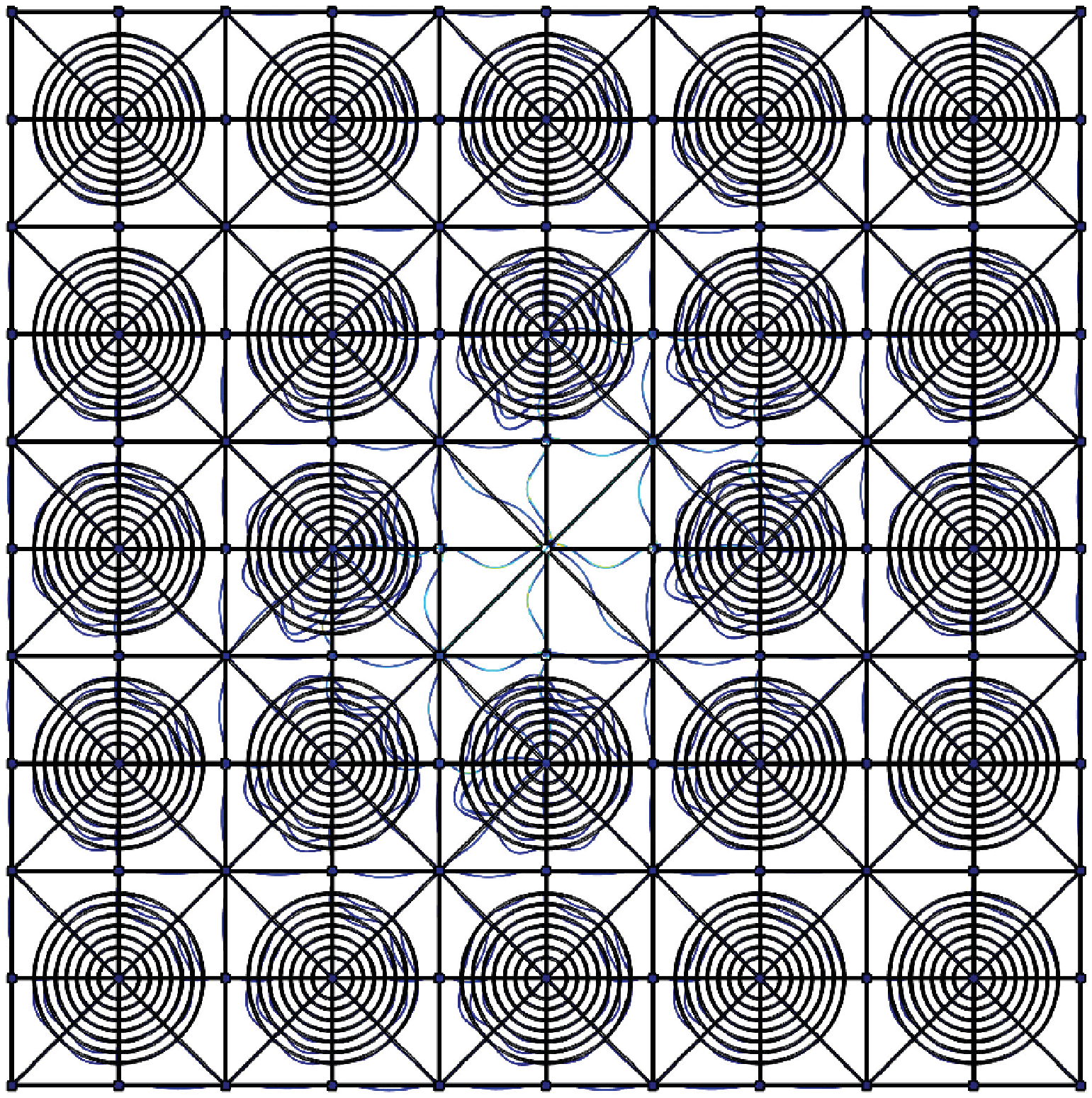}}

\caption{Frequency domain in-plane displacements (scaled by 2500) for a point excitation at an angle $\pi/4$ with respect to the  horizontal axis in (a) a regular-lattice structure and (b) spider-web inspired system with viscid-silk ring resonators. The excitation frequency 186 Hz is within a band gap for the infinite spider-web inspired metamaterial. Red and blue colors indicate maximum and minimum displacements, respectively.}
\label{FrequencyAnalysis}
\end{figure}

Having demonstrated  that infinite spider-web inspired lattice materials are characterized
by locally resonant band gaps, we now investigate how the transmission in finite-size structures 
is affected. The analysis of finite-size structure is performed in the frequency domain 
for a model comprising 25 unit cells placed in a square array with traction-free boundary conditions. 
The structure is excited at the central point by applying harmonic in-plane displacement at 
a frequency of 186 Hz (within a band gap) at an angle of $\pi/4$ with respect to the horizontal axis.
Figure~\ref{FrequencyAnalysis} presents frequency-domain responses in terms of in-plane displacements 
for two structures formed by the regular and spider-web lattice unit cells with viscid silk 
ring resonators. A scale factor of 2500 is applied to highlight the dynamics of the structure. 
Maximum and minimum values of displacements are shown in red and dark blue, respectively. 
As can be easily seen, all of the regular-lattice structure is involved in the motion (Fig.~\ref{FrequencyAnalysis}a), while the spider-web inspired system is capable of strongly attenuating vibrations after a few unit cells (Fig.~\ref{FrequencyAnalysis}b).
A similar behavior can be observed for other excitation frequencies within the predicted band gaps.
These results confirm the predictions derived from the wave dispersion analysis.
Figure~\ref{FrequencyAnalysis}b suggests an important application, namely the generation of a defect mode in a cluster with
highly localized vibrations around its center to obtain efficient wave attenuation effects in desired frequency ranges.

In summary, we have numerically studied the propagation characteristics of elastic
waves in regular and spider web-inspired beam lattices, based on the Nephila orb web architecture. 
Our results indicate
that the spider-web inspired lattices possess locally resonant band gaps 
induced by ring-shaped resonators and parts of the bearing frame.
The band gaps can be easily tuned in a wide range of frequencies
by varying the mechanical properties or the number of the resonators.
Interestingly, despite the fact that the ring resonators are 
responsible for the generation of band gaps, their eigenfrequencies cannot be directly used 
to predict the band gap bounds, since the overall structure plays an important role in their formation.
We have found that mechanical properties of the connectors between ring resonators and the lattice
frame also influence the inhibited frequency ranges. 
Though lattice systems with locally resonant band gaps have already been 
reported~\cite{WangBertoldi2015, LimBertoldi2015}, this study shows that spider-web inspired
lattice metamaterials possess more parameters to tune the band gaps to desired
frequencies and are easier to manipulate/manufacture compared to hierarchically organized lattice-type structures.
In the proposed metamaterial configuration, the role of the ring resonators is three-fold:
(1) they govern the local resonance mechanisms responsible for inducing sub-wavelength band gaps, 
(2) they contribute to the formation of band gaps due to local resonance in the bearing frame
and
(2) they are parts of the structural lattice coupled to the bearing frame to enhanced wave attenuation
performance with significant reductions of the wave energy transferred through the web architecture.
Thus, results from this study can lead to new ideas for the design of lightweight and robust bio-inspired
metamaterial structures with tunable properties. This work also suggests a new functionality for spider webs 
and new applications for the corresponding metamaterials and metastructures, e.g. for earthquake protection of suspended bridges.


\vspace{1cm}

\textbf{Acknowledgements}

M.M. has received funding from the European Union's 
Horizon 2020 research and innovation programme under the Marie 
Sk{\l}odowska-Curie grant agreement N. 658483. 
A.K. has received funding from the European Union's Seventh Framework
programme for research and innovation under the Marie 
Sk{\l}odowska-Curie grant agreement N. 609402-2020 researchers: Train to Move (T2M).
N.M.P. is supported by the European Research Council in the following active projects: ERC StG Ideas 2011 BIHSNAM no. 279985; 
 ERC PoC 2015 SILKENE no. 693670; as well as by the European Commission under the Graphene Flagship (WP ‘14 Polymer Nanocomposites’, no. 696656). F.B. is supported by BIHSNAM.


\end{document}